# Polarization and fundamental sensitivity of $^{39}$K($^{133}$Cs)-$^{85}$Rb-$^{21}$Ne co-magnetometers

Jian-Hua Liu(刘建华),[1,3] Dong-Yang Jing(靖东洋),[1,2] Lin Zhuang(庄琳),[4] Wei Quan(全伟),[5] Jiancheng Fang(房建成),[5] and Wu-Ming Liu(刘伍明)[1,2]

[1]*Beijing National Laboratory for Condensed Matter Physics, Institute of Physics, Chinese Academy of Sciences, Beijing 100190, China*
[2]*School of Physical Sciences, University of Chinese Academy of Sciences, Beijing 100190, China*
[3]*School of Science, Beijing Technology and Business University, Beijing 100048, China*
[4]*School of Physics, Sun Yat-Sen University, Guangzhou 510275, China*
[5]*School of Instrument Science and Opto-Electronics Engineering, and Science and Technology on Inertial Laboratory, Beihang University, Beijing 100191, China*

†*Corresponding author. E-mail:* wliu@iphy.ac.cn

**Abstract:** The hybrid optical pumping spin exchange relaxation free (HOPSERF) atomic co-magnetometers make ultrahigh sensitivity measurement of inertia achievable. The wall relaxation rate has a big effect on the polarization and fundamental sensitivity for the co-magnetometer, but it is often neglected in the experiments. However, there is almost no work about the systematic analysis of the influence factors on the polarization and the fundamental sensitivity of the HOPSERF co-magnetometers. Here, we systematically study the polarization and the fundamental sensitivity of $^{39}$K-$^{85}$Rb-$^{21}$Ne and $^{133}$Cs-$^{85}$Rb-$^{21}$Ne HOPSERF co-magnetometers with low polarization limit and the wall relaxation rate. The $^{21}$Ne number density, the power density and wavelength of pump beam will affect the polarization greatly by affecting the pumping rate of pump beam. We obtain a general formula on the fundamental sensitivity of the HOPSERF co-magnetometers due to shot-noise and the fundamental sensitivity changes with multiple systemic parameters, where the suitable number density of buffer gas and quench gas make the fundamental sensitivity highest. The fundamental sensitivity $7.5355\times10^{-11}$ rad/s/Hz$^{1/2}$ of $^{133}$Cs-$^{85}$Rb-$^{21}$Ne co-magnetometer is higher than the ultimate theoretical sensitivity $2\times10^{-10}$ rad/s/Hz$^{1/2}$ of K-$^{21}$Ne co-magnetometer.

## 1. Introduction

In recent years, ultrahigh sensitive co-magnetometers have become a hotspot in research of inertial navigation, geophysics,[1,2] gravitational wave detection,[3] downhole orientation sensing[4] and general relativity test[5]. Ring laser gyroscopes and fiber optic gyroscopes based on the Sagnac effect are widely used in sea and space navigation[6]. With the rapid development of quantum physics, the spin exchange relaxation free (SERF) atomic spin co-magnetometer[7] uses hyperpolarized nuclear spins to sense rotation. Atomic co-magnetometers[8,9] use two or more spin species with different gyromagnetic ratios occupying the same volume to cancel the sensitivity of the co-magnetometers to random changing magnetic field and this leaves them only sensitive to rotation or other fields. Atomic co-magnetometers are also used to search violation of local Lorentz invariance,[10,11] to study spin-dependent forces[12–15] and to search electric dipole moments[16]. A SERF atomic co-magnetometer based on K-$^{3}$He[17] reached rotation sensitivity of $5\times10^{-7}$ rad/s/Hz$^{1/2}$ in 2005. A Cs-$^{129}$Xe co-magnetometer was also studied[18]. Due to the formation of van der Waals molecules between Cs and $^{129}$Xe,[19] the Cs relaxation rate is much larger than that of the Rb in a Rb-$^{21}$Ne pair. A K-Rb-$^{21}$Ne co-magnetometer ensures high sensitivity for rotation sensing. Theoretical analysis shows that the fundamental rotation sensitivity[17] of a K-$^{21}$Ne atomic co-magnetometer could reach 2×10⁻¹⁰ rad/s/Hz^(1/2) with 10 cm³ sense volume as the density of K is 1×10¹⁴ cm⁻³, the number density of buffer gas $^{21}$Ne is 6×10¹⁹ cm⁻³. A dual-axis K-Rb-$^{21}$Ne comagnetometer can suppress the cross-talk effect and carry out high-precision rotation sensing along two sensitive axes simultaneously and independently[20] and a parametrically modulated dual-axis Cs-Rb-Ne atomic spin gyroscope can effectively suppress low frequency drift and achieve a bias instability of less than 0.05 deg/h[21]. The synchronous measurement of inertial rotation and magnetic field in a K-Rb-$^{21}$Ne comagnetometer based on the nuclear spin magnetization of the $^{21}$Ne self-compensation magnetic field and enhancement of the rotation signal[22] and the real-time closed-loop control of the compensation point[23] in a K−Rb−$^{21}$Ne comagnetometer become the focus of research for rotation sensing.

In this paper, we study the polarization and fundamental sensitivity of the HOPSERF co-magnetometer taking the wall relaxation rate into account, which has a big effect on the polarization and fundamental sensitivity, but usually it is neglected in the experiments. We obtain a general formula on the fundamental sensitivity with a low polarization limit, which describes the fundamental sensitivity of the co-magnetometer changing with the number density of buffer gas and quench gas, wavelength of pump beam, mole fraction of $^{85}$Rb, power density of pump beam, external magnetic field, cell effective radius (the shape of the cell is roughly spherical), measurement volume and cell temperature. We have investigated $^{39}$K-$^{85}$Rb-$^{21}$Ne and $^{133}$Cs-$^{85}$Rb-$^{21}$Ne HOPSERF atomic magnetometers($^{39}$K ($^{133}$Cs)-$^{85}$Rb-$^{21}$Ne co-magnetometers), then found that the fundamental sensitivity of $^{133}$Cs-$^{85}$Rb-$^{21}$Ne co-magnetometer is higher than the $^{39}$K-$^{85}$Rb-$^{21}$Ne co-magnetometer at the same cell temperature in the SERF regime with the same frequency detuning of a pump beam when the external magnetic field is smaller than about 1.7884×10⁻⁸ T, the mole fraction of $^{85}$Rb is larger than about a critical value 0.9662 or the power density of the pump beam is smaller than about 0.229 W/cm² under our chosen conditions. Optimizing the co-magnetometer parameters is advantageous to improve the sensitivity of the co-magnetometer in measuring weak rotation signal.

Furthermore, we obtain a higher fundamental sensitivity of about 7.5355×10⁻¹¹ rad/s/Hz^(1/2) with $^{133}$Cs-$^{85}$Rb-$^{21}$Ne co-magnetometer when the polarization of $^{85}$Rb atom is about 8.193×10⁻⁴, the measurement time is 1 s, cell temperature is 406.696 K, the number density of $^{85}$Rb is about 1×10¹⁴ cm⁻³, cell effective radius $a$=2 cm, measurement volume is 10 cm³, external magnetic field $B$=1×10⁻¹² T, the number density of quench gas $N_2$ is 6.3×10¹⁸ cm⁻³, the number density of buffer gas $^{21}$Ne is 6×10¹⁹ cm⁻³ and the fundamental sensitivity is higher than the fundamental rotation sensitivity[17] of a K-$^{21}$Ne atomic co-magnetometer of 2×10⁻¹⁰ rad/s/Hz^(1/2). These findings not only optimize the parameters for the SERF regime, but also provide an experimental guide for design of SERF co-magnetometers.

## 2. The principle of the HOPSERF atomic co-magnetometers

### 2.1. The number density of alkali-metal atoms

The SERF atomic co-magnetometers have some properties similar to the SERF atomic magnetometers. As we have discussed in our previous work,[24] we take the alkali metal vapor cell (the shape of the cell is roughly spherical) of the SERF atomic co-magnetometers based on HOP contain two types of alkali metal atoms, $^{39}$K-$^{85}$Rb or $^{133}$Cs-$^{85}$Rb, we take $^{39}$K or $^{133}$Cs as atom A, select $^{85}$Rb as atom B in the SERF regime,[24,25] $^{21}$Ne as buffer gas to suppress the spin relaxation and $N_2$ as quench gas to restrain radiative de-excitation of alkali metal atoms[26]. The saturated density[27] of the alkali-metal atoms vapor in units of cm⁻³ at cell temperature $T$ in Kelvin is $n_{sat} = \frac{1}{T} 10^{21.866+A_1-B_1/T}$. $A_1$ and $B_1$ are phase parameters[25], where $A_1^K$ =4.402, $A_1^{Rb}$ = 4.312, $A_1^{Cs}$ =4.165, $B_1^K$ =4453, $B_1^{Rb}$ =4040 and $B_1^{Cs}$ =3830 for the temperature higher than 400 K. Because the SERF regime can be reached by operating with sufficiently high alkali metal number density (at higher temperature) and in sufficiently low magnetic field,[28, 29] as we discussed in our previous work,[24] we choose $T$=457.5 K as the highest temperature to reduce the corrosion of alkali metal atoms to the vapor cell and make the co-magnetometers be in the SERF regime.

## 2.2. The polarization of alkali-metal atom

Considering the spin exchange between alkali-metal atoms A and B in the hybrid vapor cell, we assume that the vapor densities obey the Raoult's law,[30] $n_B \approx f_B n_{sat}^B$, where $f_B$ is the mole fraction of atom B in the metal and $n_{sat}^B$ is the saturated vapor density for pure atom B metal. When the mole fraction of atom B is 0.97, we can obtain the number density of atom A and B, $n_A \approx 0.03 n_{sat}^A$, $n_B \approx 0.97 n_{sat}^B$. For K-Rb-$^{21}$Ne atomic co-magnetometer, we suppose that in an alkali-metal vapor cell filled with K, Rb and $^{21}$Ne atoms, the density ratio of K to Rb is $D_r = n_K/n_{Rb}$ (where $n_K$ and $n_{Rb}$ are the number densities of K and Rb, respectively). K atom spins are directly polarized by the D1 line light and Rb atom spins are polarized through spin exchange with K atom spins. If the frequency rate of the laser beam $v$, is tuned away from the absorption center of the pressure[31] and the power-broadened[32] D1 absorption line, there will be an AC Stark shift of K, which is denoted as $L_z^K$ in the light propagation direction,[35] where the light propagation direction is in the $z$ direction. For Rb atom spins, the far-off-resonant laser will cause an AC Stark shift $L_z^{Rb}$ on the D1 and D2 line transitions of Rb atoms. K atom spins polarize Rb atom spins by the spin-exchange interaction. The spin transfer rate from Rb to K is given by $R_{Rb-K}^{SE} = n_{Rb} \sigma_{SE}^{Rb-K} \bar{v}_{Rb-K}$, where $\sigma_{SE}^{Rb-K}$ is the spin-exchange cross section between Rb and K[33] and $v_{Rb-K}$ is the relative velocity between Rb and K. Thus the spin transfer rate from K to Rb is $R_{K-Rb}^{SE} = D_r R_{Rb-K}^{SE}$.

The full Bloch equations are given as follows:[34, 35]

$$\frac{\partial \vec{P}_K^e}{\partial t} = \frac{\gamma_e}{Q(P_K^e)} \left(\vec{B} + \vec{L}^K\right) \times \vec{P}_K^e + \frac{R_p s_p - R_p \vec{P}_K^e + R_{Rb-K}^{SE}\left(\vec{P}_{Rb}^e - \vec{P}_K^e\right)}{Q(\vec{P}_K^e)}, \quad (1)$$

$$\frac{\partial \vec{P}_{Rb}^e}{\partial t} = \frac{\gamma_e}{Q(P_{Rb}^e)} \left(\vec{B} + \lambda_{Rb-Ne} M_0^n \vec{P}^n + \vec{L}^{Rb}\right) \times \vec{P}_{Rb}^e + \frac{D_r R_{Rb-K}^{SE}\left(\vec{P}_K^e - \vec{P}_{Rb}^e\right)}{Q(\vec{P}_{Rb}^e)},$$
$$- \frac{1}{T_{2e}, T_{2e}, T_{1e}} \frac{\vec{P}_{Rb}^e}{Q(\vec{P}_{Rb}^e)} \quad (2)$$

$$\frac{\partial \vec{P}^n}{\partial t} = \gamma^n \left(\vec{B} + \lambda_{Rb-Ne} M_0^{Rb} \vec{P}_{Rb}^e\right) \times \vec{P}^n + \vec{\Omega} \times \vec{P}^n + R_{Rb-Ne}^{se}\left(\vec{P}_{Rb}^e - \vec{P}^n\right),$$
$$- \frac{1}{T_{2n}, T_{2n}, T_{1n}} \vec{P}^n \quad (3)$$

where $\mathbf{P_K}$, $\mathbf{P_{Rb}}$ and $\mathbf{P^n}$ are the polarizations of K, Rb and $^{21}$Ne atom spins, respectively, and $\Omega$ is the input rotation velocity. $Q(P_K^e)$ and $Q(P_{Rb}^e)$ are the slow-down factors caused by the rapid spin exchange.[36] The rapid spin exchange between the K and Rb spins can change the slow-down factors of both K and Rb atom spins and the slow-down factors will be the same when they involve collision with each other.[35] $\mathbf{B}$ is the external magnetic field. $L_K$ and $L_{Rb}$ are the AC Stark shifts of the K atoms and Rb atoms respectively. $R_p$ is the pumping rate of K atoms(or the pumping rate of Cs atoms for the SERF atomic co-magnetometers based on $^{133}$Cs-$^{85}$Rb), which is mainly determined by pumping laser parameters,[17] while $\vec{s}_p$ gives the direction and magnitude of photon spin polarization. Here $\lambda_{Rb-Ne} M_0^n \vec{P}^n$ is the magnetic field produced by the $^{21}$Ne atom spins through spin-exchange interaction with Rb atom spins[37].

$\lambda_{Rb\text{-}Ne} = 8\pi\kappa_{Rb\text{-}Ne}/3$, $\kappa_{Rb-Ne}$ is a spin exchange enhancement factor resulting from the overlap of the Rb electron wave function with the $^{21}$Ne nucleus and is approximately 35.7[38]. $M_0^n = \mu_{Ne} n_{Ne}$, $\mu_{Ne}$ is nuclear magnetic moment and $n_{Ne}$ is the $^{21}$Ne atom number density. Rb atom spins also produce a magnetic field $\lambda_{Rb-Ne} M_0^{Rb} \vec{P}_{Rb}$, which is experienced by the $^{21}$Ne spins. We take $1/T_{2e}$ and $1/T_{1e}$ as the transverse and longitude relaxation rates of Rb atom spins, respectively;

$$\frac{1}{T_{2e}, T_{2e}, T_{1e}} \vec{P}_{Rb}^e = \left(\frac{1}{T_{2e}} P_{xRb}^e; \frac{1}{T_{2e}} P_{yRb}^e; \frac{1}{T_{1e}} P_{zRb}^e\right), \frac{1}{T_{2n}, T_{2n}, T_{1n}} \vec{P}^n = \left(\frac{1}{T_{2n}} P_x^n; \frac{1}{T_{2n}} P_y^n; \frac{1}{T_{1n}} P_z^n\right),$$

$1/T_{1e} = R_{SD}^{Rb} + D_r R_p + R_{Wall}^{Rb}$, $R_{SD}^{Rb} = n_{Rb}\sigma_{SD}^{Rb\text{-}Rb}\bar{v}_{Rb\text{-}Rb} + n_{Ne}\sigma_{SD}^{Rb\text{-}Ne}\bar{v}_{Rb\text{-}Ne} + n_{N_2}\sigma_{SD}^{Rb\text{-}N_2}\bar{v}_{Rb\text{-}N_2} + n_{Rb}\sigma_{SD}^{Rb\text{-}K}\bar{v}_{Rb\text{-}K}$, $R_{Wall}^{Rb}$ is the relaxation rates due to diffusion of Rb atoms to the wall (the wall relaxation of Rb atoms),[17]

$$R_{Wall}^{Rb} = Q(P_{Rb}^e) D_{Ne}^{Rb}\left(\sqrt{\frac{1+T/273.15}{P_{Ne}/1amg}}\right)\left(\frac{\pi}{a}\right)^2 + Q(P_{Rb}^e) D_{N_2}^{Rb}\left(\sqrt{\frac{1+T/273.15}{P_{N_2}/1amg}}\right)\left(\frac{\pi}{a}\right)^2,$$

$D_{Ne}^{Rb}$ is the diffusion constant of the alkali atom Rb within the buffer gas Ne[39-41] in units of cm$^2$/s and is given at 1 amg and 273 K, $D_{N_2}^{Rb}$ is the diffusion constant of the alkali atom Rb within the quench gas N$_2$ [40-42] in units of cm$^2$/s and is given at 1 amg and 273 K, 1amg = 2.69×10$^{19}$ cm$^{-3}$, $D_{Ne}^{Rb} = 0.2$ cm$^2$/s, $D_{N_2}^{Rb} = 0.19$ cm$^2$/s, $P_{Ne}$ is the pressure of buffer gas Ne in amg, $P_{N2}$ is the pressure of quench gas N$_2$ in amg, $a$ is the equivalent radius of vapor cell. The transverse relaxation rate of Rb atom spins comes from the spin-exchange relaxation rate of Rb atoms[43,44] $Q(P_{Rb}^e)/T_2^{SE}$, and $1/T_{1e}$, $1/T_{2e} = 1/T_{1e} + Q(P_{Rb}^e)/T_2^{SE}$. $R_{Rb-Ne}^{se}$ is the spin-exchange transfer rate from Rb to $^{21}$Ne atom spins. The AC Stark shift of Rb atom spins is measured by measuring the polarization of Rb atom spins along the $x$ direction $P_{xRb}^e$, with a probe laser whose wavelength is tuned to the Rb D1 line. Suppose that a $B_z$ magnetic field is applied to the co-magnetometer. The virtual magnetic fields of K and Rb caused by AC Stark shifts are in the $z$ direction. At steady state, $\partial P_K^e/\partial t = 0$, $\partial P_{Rb}^e/\partial t = 0$ and $\partial P^n/\partial t = 0$. For the low polarization limit, $R_{Wall}^{Rb} = Q(0)_{Rb} D_{Ne}^{Rb}\left(\sqrt{\frac{1+T/273.15}{P_{Ne}/1amg}}\right)\left(\frac{\pi}{a}\right)^2 + Q(0)_{Rb} D_{N_2}^{Rb}\left(\sqrt{\frac{1+T/273.15}{P_{N_2}/1amg}}\right)\left(\frac{\pi}{a}\right)^2$.

We can find that the relaxation rates for diffusion of $^{85}$Rb in the $^{21}$Ne gas to the wall of $^{39}$K-$^{85}$Rb-$^{21}$Ne co-magnetometer $R_{wall-K-Rb-Ne}^{Rb-Ne}$, the total relaxation rates for diffusion of $^{85}$Rb in the $^{21}$Ne and N$_2$ gas to the wall of $^{39}$K-$^{85}$Rb-$^{21}$Ne co-magnetometer $R_{wall-K-Rb-Ne}^{Rb-Ne-N_2}$, the relaxation rates for diffusion of $^{85}$Rb in the $^{21}$Ne gas to the wall of $^{133}$Cs-$^{85}$Rb-$^{21}$Ne co-magnetometer $R_{wall-Cs-Rb-Ne}^{Rb-Ne}$ and the total relaxation rates for diffusion of $^{85}$Rb in the $^{21}$Ne and N$_2$ gas to the wall of $^{133}$Cs-$^{85}$Rb-$^{21}$Ne co-magnetometer $R_{wall-Cs-Rb-Ne}^{Rb-Ne-N_2}$ decrease when $^{21}$Ne atom number density $n_{Ne}$ increases in Fig. 1(a), because the buffer gas $^{21}$Ne atom suppresses the spin relaxation caused by wall collisions. The relaxation rates for diffusion of $^{85}$Rb in the N$_2$ gas to the wall of $^{39}$K-$^{85}$Rb-$^{21}$Ne co-magnetometer is $R_{wall-K-Rb-Ne}^{Rb-N_2}$, the relaxation rates for diffusion of $^{85}$Rb in the N$_2$ gas to the wall of $^{133}$Cs-$^{85}$Rb-$^{21}$Ne co-magnetometer is $R_{wall-Cs-Rb-Ne}^{Rb-N_2}$. $R_{wall-K-Rb-Ne}^{Rb-N_2}$, $R_{wall-K-Rb-Ne}^{Rb-Ne-N_2}$, $R_{wall-Cs-Rb-Ne}^{Rb-N_2}$ and $R_{wall-Cs-Rb-Ne}^{Rb-Ne-N_2}$ decrease when $n_{N2}$ increases for N$_2$ gas colliding with excited alkali metal atoms and absorbing the energy, which restrains radiative de-excitation of the excited alkali metal atoms, $R_{wall-K-Rb-Ne}^{Rb-Ne}$ and $R_{wall-Cs-Rb-Ne}^{Rb-Ne}$ do not change when $n_{N2}$ increases because they are not related with $n_{N2}$ in Fig. 1(b).

The $R_{wall-K-Rb-Ne}^{Rb-N_2}$, $R_{wall-K-Rb-Ne}^{Rb-Ne-N_2}$, $R_{wall-Cs-Rb-Ne}^{Rb-N_2}$, $R_{wall-Cs-Rb-Ne}^{Rb-Ne-N_2}$, $R_{wall-K-Rb-Ne}^{Rb-Ne}$ and $R_{wall-Cs-Rb-Ne}^{Rb-Ne}$ increase with increasing cell temperature $T$ in Fig. 1(c) because the increasing $T$ makes the number density and relative velocity of $^{85}$Rb atoms greater and more $^{85}$Rb atoms move towards the wall. The $R_{wall}$ decreases when the cell effective radius $a$ increases in Fig. 1(d) for the $^{85}$Rb atoms colliding more $^{21}$Ne atoms and N$_2$ atoms with the increasing $a$ when $^{85}$Rb atoms move towards the wall, which will decrease the wall relaxation rate.

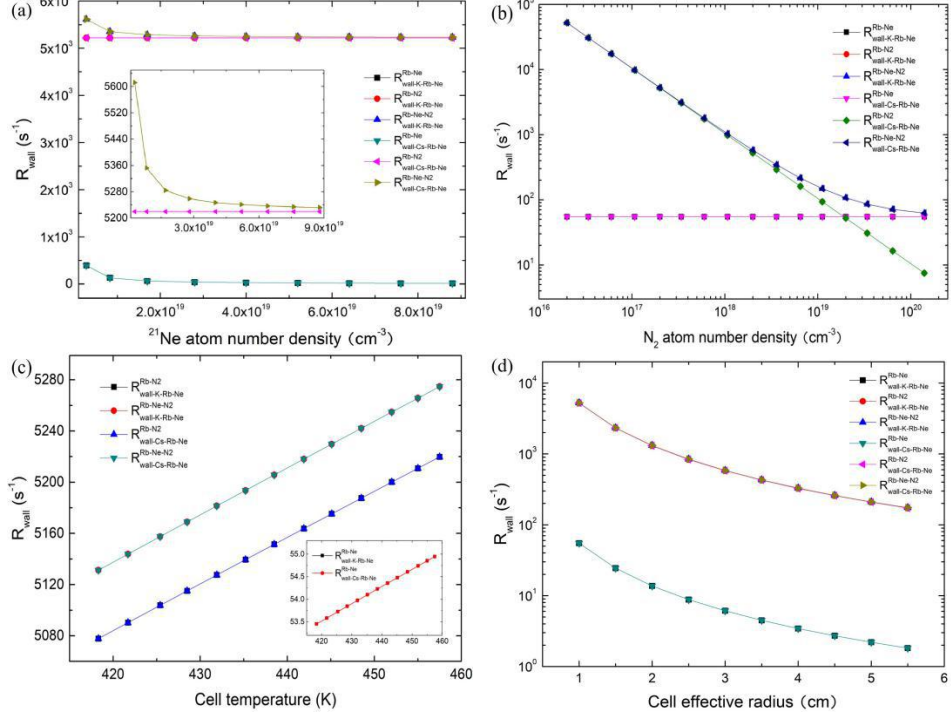

Fig. 1. The wall relaxation rates change with the $^{21}$Ne number density $n_{Ne}$ in (a), the N$_2$ number density $n_{N2}$ in (b), the cell temperature $T$ in (c) and the cell effective radius $a$ in (d), respectively.

The polarization in the $z$ direction could be calculated by[35]

$$P_{zK}^e \approx P_{zRb}^e \approx \frac{D_r R_p}{D_r R_p + R_{SD}^{Rb} + Q(0)_{Rb}/T_{1e}} \quad (\mathrm{R_p} \ll \mathrm{R_{Rb-K}^{SE}},\ \mathrm{D_r} \ll 1),\quad P_z^n \approx \frac{R_{Rb-Ne}^{se}}{R_{Rb-Ne}^{se} + 1/T_{1n}}.$$

The angle between the spins polarization and the $z$ direction is small by applying a small magnetic field $B_y$ and the polarizations of K, Rb and $^{21}$Ne are approximately constant in the $z$ direction, then the coupled Bloch equations can be simplified.[35] In the $x$ direction, the polarizations of K and Rb can be solved and are given by[35]

$$P_{xK}^e \approx P_{xRb}^e \approx P_{zRb}^e \frac{\left(R_{tot}^{Rb}/\gamma_e\right)\left(\frac{B_y}{B^n}\delta B_z + \frac{\Omega_y}{\gamma^n} + \frac{L_z}{R_{tot}^{Rb}/\gamma_e}\frac{\Omega_x}{\gamma^n}\right)}{\left(R_{tot}^{Rb}/\gamma_e\right)^2 + \left(\delta B_z + L_z\right)^2}, \qquad (4)$$

where $P_{zRb}^e \approx \dfrac{D_r R_p}{D_r R_p + R_{SD}^{Rb} + Q(0)_{Rb}/T_{1e}}$, $\dfrac{1}{T_{1e}} = D_r R_p + R_{SD}^{Rb} + R_{Rb}^{Wall}$,

$R_{tot}^{Rb} = D_r R_p + R_{SD}^{Rb} + R_{Rb}^{Wall} + Q(0)_{Rb}/T_2^{SE}$, $R_p = R_p^K$ or $R_p^{Cs}$, $R_p^K = \dfrac{\Phi_{D1}^K r_e c f_{D1}^K}{A_t}\dfrac{\Gamma_{D1}^K/2}{(\nu_K - \nu_{0D1}^K)^2 + (\Gamma_{D1}^K/2)^2}$,

$R_p^{Cs} = \dfrac{\Phi_{D1}^{Cs} r_e c f_{D1}^{Cs}}{A_t}\dfrac{\Gamma_{D1}^{Cs}/2}{(\nu_{Cs} - \nu_{0D1}^{Cs})^2 + (\Gamma_{D1}^{Cs}/2)^2}$, $\delta B_z = B_z - B^e - B^n$, $B^e \approx \dfrac{8}{3}\pi\mu_0\kappa_{Rb-Ne}M_0^{Rb}P_{zRb}^e$,

$$M_0^{Rb} \approx \mu_B n_{Rb}, \quad B^n \approx \frac{8}{3}\pi\mu_0 \kappa_{Rb-Ne}\mu_{Ne}n_{Ne}\frac{R_{Rb-Ne}^{se}}{R_{Rb-Ne}^{se}+1/T_{1n}}, \quad T_{1n} \approx 214 \times 60 \times 2.69 \times 10^{19}/n_{Ne},$$

$$\mu_0 = 4\pi \times 10^{-7} \text{Tm/A}, \quad \mu_{Ne} = 3.33 \times 10^{-27} \text{J/T}, \quad R_{Rb-Ne}^{se} = \kappa_{Rb-Ne}^{se} n_{Rb}, \quad \kappa_{Rb-Ne}^{se} = \sigma_{SD}^{Rb-Ne}\bar{v}_{Rb-Ne},$$

$$L_z = D_r L_z^K + L_z^{Rb}, \quad L_z^K = \frac{\Phi_{D1}^K r_e c f_{D1}^K}{A_t \gamma_e} \frac{\nu_K - \nu_{0D1}^K}{(\nu_K - \nu_{0D1}^K)^2 + (\Gamma_{D1}^K/2)^2}, \quad \Gamma_{D1}^K = \Gamma_{P-D1}^K \sqrt{1+\frac{3\lambda^3 I_{PI}^K}{\pi h c \Gamma_{P-D1}^K}},$$

$$\Gamma_{P-D1}^K = \frac{n_{Ne}P_0 T}{T_0 \times 2.69 \times 10^{19}/cm^3} \times 7.8 GHz/atm, \quad L_z^{Rb} = \frac{\Phi_{D1}^K r_e c f_{D1}^{Rb}}{A_t \gamma_e}\left(\frac{1}{\nu_K - \nu_{0D2}^{Rb}} - \frac{1}{\nu_K - \nu_{0D1}^{Rb}}\right), \quad T_0 = 273.15K,$$

$P_0$=101.325kPa=1atm. $1/T_{2n}$ and $1/T_{1n}$ are the transverse and longitude relaxation rates[38] of the $^{21}$Ne atom spins, respectively. We can obtain the polarizations of Cs and Rb in the $x$ direction for $^{133}$Cs-$^{85}$Rb-$^{21}$Ne co-magnetometers by replacing the K atoms by Cs atoms in Eq. (4). The linewidth of the Cs-Ne pressure broadening $\Gamma_{P-D1}^{Cs} = \frac{n_{Ne}P_0 T}{T_0 \times 2.69 \times 10^{19}/cm^3} \times 8.2GHz/atm$, $\kappa_{Rb-Ne} \approx 35.7$, $I_{PI}^K$ ($I_{PI}^{Cs}$) is the power density of pump beam for $^{39}$K ($^{133}$Cs)-$^{85}$Rb-$^{21}$Ne co-magnetometers, the $^{21}$Ne relaxation is dominated by the electric quadrupole interaction. During binary collisions, the interaction between the induced electric-field gradients and the nuclear quadrupole moments produce torques on the spins, thus cause relaxation[46]. The resonance transition of $\lambda_{0D1}^K$ is 770.108 nm, $\nu_{0D1}^K = c/\lambda_{0D1}^K \approx 3.895 \times 10^5$ GHz, the D1 transitions of Rb and Cs in vacuum are $\lambda_{0D1}^{Rb}$ =795 nm and $\lambda_{0D1}^{Cs}$ =894.6 nm, respectively. Therefore, $\nu_{0D1}^{Rb} = c/\lambda_{0D1}^{Rb} \approx 3.773 \times 10^5$ GHz, $\nu_{0D1}^{Cs} = c/\lambda_{0D1}^{Cs} \approx 3.353 \times 10^5$ GHz, the D2 transition of K and Rb in vacuum are $\lambda_{0D2}^K$ =766.7nm and $\lambda_{0D2}^{Rb}$ =780.2 nm, respectively, $\nu_{0D2}^{Rb} = c/\lambda_{0D2}^{Rb} \approx 3.845 \times 10^5$ GHz. $f_{D1}^K = 0.324$, $f_{D1}^{Cs} = 0.347$, $f_{D1}^{Rb} = 0.332$, $f_{D2}^{Rb} = 0.668$.[47] $\Phi_{D1}^K$ is the photon number flux, $\Phi_{D1}^K = \lambda_K I_{PI}^K \pi a_p^2/hc$ and $A_t$ is the transverse area of the pumping light, $A_t = \pi a_p^2$, $r_e$ is the classical electron radius, $r_e = 2.817938 \times 10^{-15}$ m, and $c$ is the velocity of light, $f_{D1}$ is the oscillator strength[48] and $\gamma_e$ is the electron gyromagnetic ratio, $\Gamma_{D1}^K$ is the linewidth (full width at half maximum) of the pressure and power-broadened K D1 absorption line[49], and $\nu_{0D1}^K$ is the absorption center frequency rate of the absorption line. The far-off-resonant pump laser will cause an AC Stark shift of Rb atoms. $\delta B_z$ is the compensation magnetic field.

For small atomic polarization, the spin precession rate is given by $\omega_0 = g\mu_B B/Q(P^e)\hbar$, which is the Larmor frequency, $T_{SE}$ is the spin-exchange time, $1/T_{SE}^{Rb} = n_{Rb}\sigma_{SE}^{Rb-Rb}\bar{v}_{Rb-Rb}$, where $\sigma_{SE}^{Rb-Rb}$ is the spin-exchange cross section between Rb and Rb, and $\bar{v}_{Rb-Rb}$ is the relative velocity between Rb and Rb. $Q(P^e)$ is the slow-down factor for polarization and nuclear angular momentum[27,50], $Q(P^e)_{85_{Rb}} = (38+52P^2+6P^4)/(3+10P^2+3P^4)$ for the $^{85}$Rb atom. In this limit spin exchange contributes to transverse relaxation only in second order and vanishes for zero magnetic field,[17] $1/T_2^{SE} = \omega_0^2 T_{SE}\left[Q(P^e)^2 - (2I+1)^2\right]Q(P^e)^2/2Q(P^e)^2$, for low polarization limit and small magnetic fields, spin-exchange relaxation is quadratic in the magnetic field, $I$ is nuclear spin of the alkali-metal atoms, we consider the relaxation rate due to $^{85}$Rb-$^{85}$Rb, $^{85}$Rb-$^{39}$K spin-exchange collisions, then

$$\frac{1}{T_2^{SE}} = \left(\frac{g\mu_B B}{Q(0)_{Rb}\hbar}\right)^2 \left[\frac{Q(0)_{Rb}^2 - (2I_{Rb}+1)^2}{2n_{Rb}\sigma_{SE}^{Rb-Rb}\bar{v}_{Rb-Rb}} + \frac{Q(0)_{Rb}^2 - (2I_{Rb}+1)^2}{2n_{Rb}\sigma_{SE}^{Rb-K}\bar{v}_{Rb-K}}\right], \quad (5)$$

For $^{39}$K, $^{85}$Rb and $^{133}$Cs, $I_K=3/2$, $I_{Rb}=5/2$, $I_{Cs}=7/2$, the relaxation rate for alkali-alkali spin-exchange collisions reads

$$R_{SE}^{RbRb} = \left(\frac{g\mu_B B}{Q(0)_{Rb}\hbar}\right)^2 \frac{Q(0)_{Rb}^2 - (2I_{Rb}+1)^2}{2n_{Rb}\sigma_{SE}^{Rb-Rb}\bar{v}_{Rb-Rb}}, \qquad (6)$$

$$R_{SE}^{KRb} = \left(\frac{g\mu_B B}{Q(0)_K\hbar}\right)^2 \frac{Q(0)_K^2 - (2I_K+1)^2}{2n_K\sigma_{SE}^{K-Rb}\bar{v}_{K-Rb}}, \qquad (7)$$

$$R_{SE}^{RbK} = \left(\frac{g\mu_B B}{Q(0)_{Rb}\hbar}\right)^2 \frac{Q(0)_{Rb}^2 - (2I_{Rb}+1)^2}{2n_{Rb}\sigma_{SE}^{Rb-K}\bar{v}_{Rb-K}}, \qquad (8)$$

$$R_{SE}^{KK} = \left(\frac{g\mu_B B}{Q(0)_K\hbar}\right)^2 \frac{Q(0)_K^2 - (2I_K+1)^2}{2n_K\sigma_{SE}^{K-K}\bar{v}_{K-K}}, \qquad (9)$$

where $B$ is the external magnetic field. With sufficiently high alkali metal number density (at higher temperature) and in sufficiently low magnetic field, $R_{SE}^{KK} \approx R_{SE}^{RbRb} \approx R_{SE}^{RbK} \approx R_{SE}^{KRb} \approx 0$. In the absence of spin-exchange relaxation, spin destruction collisions due to the spin-rotation interaction[46] become a limiting factor. For low polarization limit, we take $Q(P_K^e) \approx Q(0)_K$, $Q(P_{Rb}^e) \approx Q(0)_{Rb}$, $\sigma_{SE}^{K-Rb}$ is the spin-exchange cross section of $^{39}$K and $^{85}$Rb by spin-exchange collisions with each other, the values of relevant parameters are given in Table 1.

### 2.3. The fundamental sensitivity of the HOPSERF co-magnetometer

To improve the practicability of the HOPSERF co-magnetometers, it is necessary for us to investigate the fundamental sensitivity of the co-magnetometers to improve the sensitivity, stability of the co-magnetometers and to realize the miniaturization of the co-magnetometers. The fundamental shot-noise-limited sensitivity of an atomic gyroscope based on the co-magnetometer is given by[17,59]

$$\delta\Omega = \frac{\gamma^n}{\gamma_e\sqrt{nT_2 Vt}}, \qquad (10)$$

where $\gamma^n$ is the nuclear gyromagnetic ratio of the noble gas atom, $\gamma_e = g\mu_B/\hbar$,[27, 59] $g$ is the electron g-factor, $\mu_B$ is the Bohr magneton, $n$ is the density of alkali metal atoms, $V$ is the measurement volume, $t$ is the measurement time, $T_2$ is the transverse spin relaxation time, $1/T_2 = R_{wall} + R_{SD} + Q(P^e)R_{SE}^{ee}$. For the transverse spin relaxation time of the co-magnetometers, we need consider the spin destruction relaxation $R_{SD}$ caused by Ne, N$_2$, alkali metal atom A and B, the relaxation rates due to diffusion of alkali metal atoms A and B to the wall[17], i.e., $R_{wall}^A$ and $R_{wall}^B$, the relaxation rate due to alkali-alkali spin exchange collisions, i.e., $R_{SE}^{ee}$,[60] $R_{SE}^{ee} = R_{SE}^{AA} + R_{SE}^{BB} + R_{SE}^{AB} + R_{SE}^{BA}$, which cannot be ignored for large external magnetic field $B$ and is negligible in SERF regime (when $T$ is higher than 418.3 K, $B$ is smaller than $10^{-10}$ T, $R_{SE}^{ee} \approx 0$), the pumping rate of pump beam $R_p$ and the pumping rate of atom B $R_B$ ($R_B$ is a function of $R_p$), hence $1/T_2 = R_{SD}^A + R_{SD}^B + R_{wall}^A + R_{wall}^B + R_p + Q(P^e)R_{SE}^{ee} + R_B$, we substitute this term into Eq. (10) and obtain

$$\delta\Omega = \frac{\gamma^n\sqrt{R_{Wall} + R_{SD}^{AB} + Q(P^e)R_{SE}^{ee} + R_p + R_B}}{\gamma_e\sqrt{nVt}}, \qquad (11)$$

where $R_{SD}^{AB} = R_{SD}^{A} + R_{SD}^{B}$, $R_{Wall} = R_{wall}^{A} + R_{wall}^{B}$. However, because alkali metal atom B is the probed atom, only these items associated with atom B will be considered in the experiments, we do not consider those items irrelevant to atom B, for the low polarization limit, we acquire the fundamental sensitivity of the co-magnetometers due to the shot-noise as follows:

$$\delta\Omega' = \frac{\gamma^n \sqrt{R_{wall}^B + R_{SD} + R_B + Q(0)_B R_{SE}^{EE}}}{\gamma_e \sqrt{n_B V t}}, \quad (12)$$

where $R_{wall}^B = Q(0)_B D_{buffer}^B \frac{\sqrt{1+T/273.15}}{P_{buffer}/1\mathrm{amg}} (\frac{\pi}{a})^2 + Q(0)_B D_{quench}^B \frac{\sqrt{1+T/273.15}}{P_{quench}/1\mathrm{amg}} (\frac{\pi}{a})^2$, $D_{buffer}$ and $D_{quench}$ are the diffusion coefficients of the alkali atom within Ne and N$_2$ in units of cm$^2$/s and is given at 1 amg and 273 K respectively, 1 amg = 2.69×10$^{19}$ cm$^{-3}$, $P_{buffer}$ is the pressure of buffer gas in amg, $P_{quench}$ is the pressure of quench gas in amg, $a$ is the equivalent radius of vapor cell, $R_{SD} = n_B \sigma_{SD}^{B-B} \bar{v}_{B-B} + n_{Ne} \sigma_{SD}^{B-Ne} \bar{v}_{B-Ne} + n_{N_2} \sigma_{SD}^{B-N_2} \bar{v}_{B-N_2} + n_B \sigma_{SD}^{B-A} \bar{v}_{B-A} + n_A \sigma_{SD}^{A-B} \bar{v}_{A-B}$,

$R_B = D_r R_p$, $R_{SE}^{EE} = R_{SE}^{BB} + R_{SE}^{BA}$, $R_{SE}^{BB} = \left(\frac{g\mu_B B}{Q(0)_B \hbar}\right)^2 \frac{Q(0)_B^2 - (2I_B+1)^2}{2n_B \sigma_{SE}^{B-B} \bar{v}_{B-B}}$, $R_{SE}^{BA} = \left(\frac{g\mu_B B}{Q(0)_B \hbar}\right)^2 \frac{Q(0)_B^2 - (2I_B+1)^2}{2n_B \sigma_{SE}^{B-A} \bar{v}_{B-A}}$.

In the limit of fast spin-exchange and small magnetic field, the spin-exchange relaxation rate vanishes for sufficiently small magnetic field.[36] In Eq. (12), we can find that the fundamental sensitivity of the co-magnetometers increases when part or all of $R_{wall}^B$, $R_{SD}$, $R_B$, $R_{SE}^{BB}$ and $R_{SE}^{BA}$ (the later two terms are approximately zero in sufficiently low magnetic field and the co-magnetometer is in the SERF regime, which is helpful for us to study how $B$ influence the SERF regime and fundamental sensitivity of the co-magnetometers) decrease, $n_B$ and $V$ increase. For the expressions of $R_{wall}^B$, $R_{SD}$, $R_B$, $R_{SE}^{BB}$, $R_{SE}^{BA}$, $n_B$ and $V$, we just need to consider the fundamental sensitivity of the co-magnetometers change with one of the cell effective radius $a$, $n_{Ne}$, $n_{N2}$, cell temperature $T$, wavelength of pump beam ($\lambda_K$ and $\lambda_{Cs}$), power density of pump beam ($I_{PI}^K$ and $I_{PI}^{Cs}$), pump beam spot radius $a_p$, the mole fraction of $^{85}$Rb $f_{Rb}$, external magnetic field $B$ and measurement volume $V$.

## 3. Results and discussion

### 3.1. The calculation details of the polarization and fundamental sensitivity

Because the slow-down factors are different in the polarization of $^{39}$K-$^{85}$Rb-$^{21}$Ne and $^{133}$Cs-$^{85}$Rb-$^{21}$Ne co-magnetometers, we take the slow-down factors at a low polarization limit for convenience of the theoretical analysis. The number density of the $^{21}$Ne, the power density of pump beam and the pump beam wavelength will affect the polarization greatly by affecting the pumping rate of pump beam. We systematically studied the variations of the pumping rate of pump beam, frequency shift, $B^n$, $B^e$, the relaxation rate for alkali-alkali spin-exchange collisions. We obtain the following results by MATLAB and take several points to plot with Origin 8.

### 3.2. The calculation result of the polarization

We choose one of $^{21}$Ne number density $n_{Ne}$, N$_2$ number density $n_{N2}$, cell temperature $T$, wavelength of pump beam $\lambda_K(\lambda_{Cs})$, mole fraction of $^{85}$Rb $f_{Rb}$, cell effective radius $a$, and power density of pump beam $I_{PI}^K$ ($I_{PI}^{Cs}$), the input rotation velocity in the $x$ direction $\Omega_x$ and $y$ direction $\Omega_y$, external magnetic field in the $y$ direction $B_y$ and $z$ direction $B_z$ by Eq. (4) as a variable (other parameters are invariable) to obtain the results that the polarization of the $^{39}$K($^{133}$Cs)-$^{85}$Rb-$^{21}$Ne HOPSERF co-magnetometer changes with the variable. From the formula of $R_p$, we can find that $R_p$, the frequency shift and $B^e$ do not change with the increasing $a_p$ duo to $a_p$ in $\Phi_{D1}^K$ ($\Phi_{D1}^{Cs}$) and $A_t$ are canceled out. Depending on suggestions and the typical conditions of the experiment group,[34,35,54] in order to facilitate the theoretical analysis, we take the mole fraction of $^{85}$Rb atom as $f_{Rb}$ = 0.97, $n_{Ne}$ = 2×10$^{19}$ cm$^{-3}$, $n_{N2}$ = 2×10$^{17}$ cm$^{-3}$, $I_{PI}^{K} = I_{PI}^{Cs}$ =0.06 W/cm$^2$, $\Omega_x$ =7.292×10$^{-6}$ rad/s, $\Omega_y$ =7.292×10$^{-5}$ rad/s, $B_y$=1×10$^{-12}$ T, $B_z$=7×10$^{-11}$ T, $a$=1 cm, $a_p$=1 cm, $\lambda_K$= 769.808nm, $\lambda_{Cs}$=894.3nm and $T$=457.5 K, at the moment, $^{39}$K, $^{85}$Rb and $^{133}$Cs are in the SERF regime.

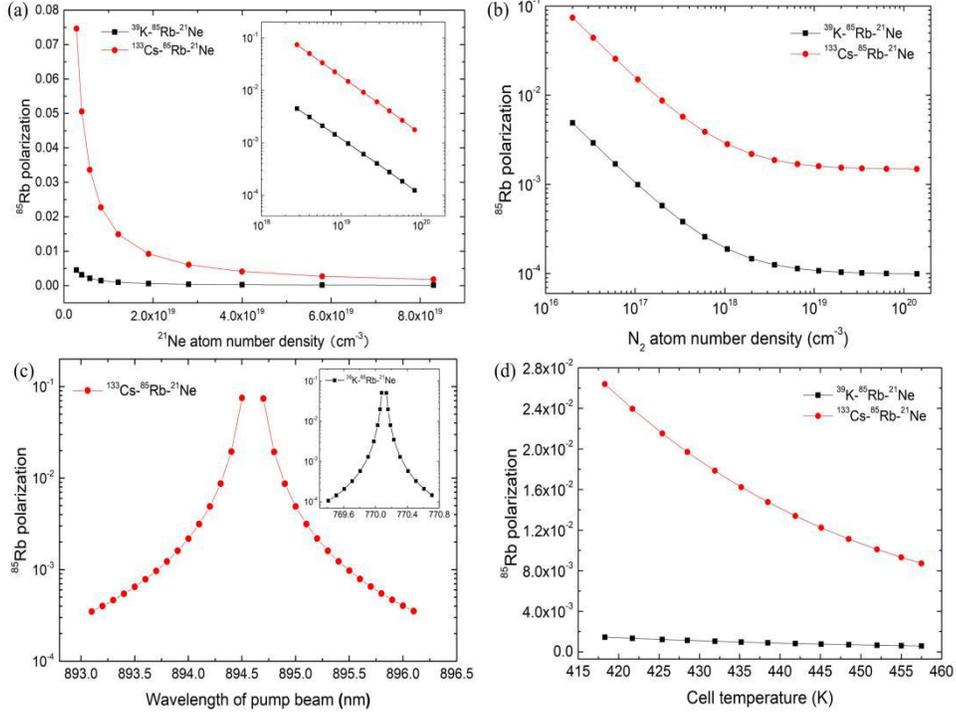

Fig.2. The $^{85}$Rb polarization of $^{39}$K ($^{133}$Cs)-$^{85}$Rb-$^{21}$Ne co-magnetometers versus the number density of the buffer gas $^{21}$Ne in (a), the number density of the quench gas N$_2$ in (b), the wavelength of pump beam in (c) and the cell temperature in (d).

To ensure the validity of the low polarization limit and considering that the noble gas has enough magnetic moment which can compensate for the external magnetic field and the system can be a co-magnetometer, we make the $^{85}$Rb polarization smaller than about 0.08 and bigger than 10$^{-4}$. We discuss the variations of the pumping rate of pump beam, the frequency shift, the $B^n$, the $B^e$, the alkali-alkali spin-exchange collisions relaxation rate related with $^{85}$Rb atom. Fig. 2 demonstrates the $^{85}$Rb polarization of $^{39}$K ($^{133}$Cs)-$^{85}$Rb-$^{21}$Ne co-magnetometers changes with $n_{Ne}$, $n_{N2}$, $\lambda_K(\lambda_{Cs})$ and $T$, respectively.

Figure 2(a) shows that $^{85}$Rb polarizations decrease with the increasing $n_{Ne}$ in $^{39}$K-$^{85}$Rb-$^{21}$Ne and $^{133}$Cs-$^{85}$Rb-$^{21}$Ne co-magnetometers. The $^{85}$Rb polarization of $^{133}$Cs-$^{85}$Rb-$^{21}$Ne co-magnetometer is larger than the one of $^{39}$K-$^{85}$Rb-$^{21}$Ne co-magnetometer.

From the formula of $R_{SD}^{Rb}$, $R_p$, T$_{1n}$, frequency shift, $B^n$, and $B^e$, we can find that $R_{SD}^{Rb}$ increases with the increasing $n_{Ne}$, $R_{wall-K-Rb-Ne}^{Rb-Ne}$, $R_{wall-Cs-Rb-Ne}^{Rb-Ne}$, $R_{wall-K-Rb-Ne}^{Rb-Ne-N_2}$, $R_{wall-Cs-Rb-Ne}^{Rb-Ne-N_2}$ decrease when $^{21}$Ne atom number density $n_{Ne}$ increases in Fig. 1(a), $R_p$, $B^n$ and $B^e$ increase with increasing $n_{Ne}$, T$_{1n}$, $L_{K-Rb-Ne}^{K}$, $L_{K-Rb-Ne}^{K-Rb}$, $L_{Cs-Rb-Ne}^{Cs}$ and $L_{Cs-Rb-Ne}^{Cs-Rb}$ decrease with increasing $n_{Ne}$. The change value of $^{85}$Rb polarization, which is mainly determined by the $R_{SD}^{Rb}$, $R_{wall}^{Rb}$ and $R_p$.

Figure 2(b) demonstrates that $^{85}$Rb polarization decreases with increasing $n_{N2}$. From the formula of $R_{SD}^{Rb}$, $R_{Wall}$ and $B^e$, we can find that $R_{SD}^{Rb}$ increases with the increasing $n_{N2}$,

$R_{wall-K-Rb-Ne}^{Rb-N_2}$, $R_{wall-Cs-Rb-Ne}^{Rb-N_2}$, $R_{wall-K-Rb-Ne}^{Rb-Ne-N_2}$ and $R_{wall-Cs-Rb-Ne}^{Rb-Ne-N_2}$ decrease when $n_{N2}$ increases, $R_{wall-K-Rb-Ne}^{Rb-Ne}$ and $R_{wall-Cs-Rb-Ne}^{Rb-Ne}$ do not change when $n_{N2}$ increases in Fig. 1(b), $B^e$ decreases slowly with increasing $n_{N2}$. Here, $^{85}$Rb polarization is mainly determined by the $R_{SD}^{Rb}$ and $R_{wall}^{Rb}$.

Figure 2(c) shows that $^{85}$Rb polarization of $^{39}$K($^{133}$Cs)-$^{85}$Rb-$^{21}$Ne co-magnetometer increases when $\lambda_K(\lambda_{Cs})$ increases but is smaller than the value of about 770.085 nm (894.5032 nm), decreases when $\lambda_K(\lambda_{Cs})$ is larger than the value of about 770.1309 nm (894.696 nm)($^{85}$Rb polarization is about 0.08). From the formulae of the $^{85}$Rb polarization, $R_p$, $L_z$ and $B^e$, we can find that $R_p$, $L_{K-Rb-Ne}^{K}$ ($L_{Cs-Rb-Ne}^{Cs}$) and $B^e$ increase and $L_{K-Rb-Ne}^{Rb}$ ($L_{Cs-Rb-Ne}^{Rb}$) decreases with increasing $\lambda_K(\lambda_{Cs})$ when the $\lambda_K(\lambda_{Cs})$ is smaller than the value of about 770.085 nm (894.5032 nm). $R_p$ and $B^e$ decrease with increasing $\lambda_K(\lambda_{Cs})$ when the $\lambda_K(\lambda_{Cs})$ is larger than the value of about 770.1309 nm (894.696 nm). Therefore, the change of $^{85}$Rb polarization with $\lambda_K(\lambda_{Cs})$ is mainly determined by the pumping rate of the pump beam, the frequency shift and $B^e$.

Figure 2(d) demonstrates that $^{85}$Rb polarization decreases with the increasing $T$. The $^{85}$Rb polarization of $^{133}$Cs-$^{85}$Rb-$^{21}$Ne co-magnetometer is larger than the one of $^{39}$K-$^{85}$Rb-$^{21}$Ne co-magnetometer. From the formula of the $R_{SD}^{Rb}$ and Fig. 1(c), $R_p$, frequency shift, $B^n$, $B^e$, alkali-alkali spin exchange collision relaxation rate, we can find that $R_{SD}^{Rb}$ increases with increasing $T$. $R_{wall-K-Rb-Ne}^{Rb-N_2}$, $R_{wall-Cs-Rb-Ne}^{Rb-N_2}$, $R_{wall-K-Rb-Ne}^{Rb-Ne-N_2}$, $R_{wall-Cs-Rb-Ne}^{Rb-Ne-N_2}$, $R_{wall-K-Rb-Ne}^{Rb-Ne}$ and $R_{wall-Cs-Rb-Ne}^{Rb-Ne}$ increase with increasing $T$ in Fig. 1(c) because the increasing $T$ makes the number density and relative velocity of $^{85}$Rb atoms larger and more $^{85}$Rb atoms move towards the wall. $R_p^K$ and $R_p^{Cs}$ increase with increasing $T$, duo to the fact that the increasing $T$ increases the linewidth of the K-Ne and Cs-Ne pressure broadening, which make $R_p^K$ and $R_p^{Cs}$ increase. $L_{K-Rb-Ne}^{K}$, $L_{Cs-Rb-Ne}^{Cs}$, $L_{K-Rb-Ne}^{K-Rb}$ and $L_{Cs-Rb-Ne}^{Cs-Rb}$, decrease slowly with increasing $T$, duo to the fact that the increasing $T$ increases the linewidth of the K-Ne and Cs-Ne pressure broadening, which make $L_{K-Rb-Ne}^{K}$, $L_{Cs-Rb-Ne}^{Cs}$, $L_{K-Rb-Ne}^{K-Rb}$ and $L_{Cs-Rb-Ne}^{Cs-Rb}$ decrease. However, $L_{K-Rb-Ne}^{Rb}$ and $L_{Cs-Rb-Ne}^{Rb}$ do not change with the $T$ because $L_{K-Rb-Ne}^{Rb}$ and $L_{Cs-Rb-Ne}^{Rb}$ have nothing to do with the $T$. $B^n$ increases with increasing $T$ duo to the fact that $R_{Rb-Ne}^{se}$ changes with $T$ from the equation of $B^n$ and $R_{Rb-Ne}^{se}$ increases with the increasing $T$. $B^e$ increases with increasing $T$. $R_{KRbNe}^{RbK}$, $R_{CsRbNe}^{RbCs}$, $R_{KRbNe}^{RbRb}$, $R_{CsRbNe}^{RbRb}$, $R_{KRbNe}^{EE}$, $R_{CsRbNe}^{EE}$ decrease with the increasing $T$ because the increasing $T$ makes $R_{Rb-K}^{SE}$, $R_{Rb-Rb}^{SE}$ and $R_{Rb-Cs}^{SE}$ increase, which is inversely proportional to the $R_{KRbNe}^{RbK}$, $R_{KRbNe}^{RbRb}$, $R_{CsRbNe}^{RbCs}$, $R_{CsRbNe}^{RbRb}$ in Eq. (5), respectively. Therefore, the total effect of the $R_{SD}^{Rb}$, the pumping rate of the pump beam ($R_p^K$ and $R_p^{Cs}$), the frequency shift ($L_{K-Rb-Ne}^{K}$, $L_{Cs-Rb-Ne}^{Cs}$, $L_{K-Rb-Ne}^{K-Rb}$ and $L_{Cs-Rb-Ne}^{Cs-Rb}$), $B^n$, $B^e$ and the alkali-alkali spin-exchange collisions relaxation rate ($R_{KRbNe}^{RbK}$, $R_{CsRbNe}^{RbCs}$, $R_{KRbNe}^{RbRb}$, $R_{CsRbNe}^{RbRb}$, $R_{KRbNe}^{EE}$, $R_{CsRbNe}^{EE}$) make the $^{85}$Rb polarization decrease with the increasing $T$.

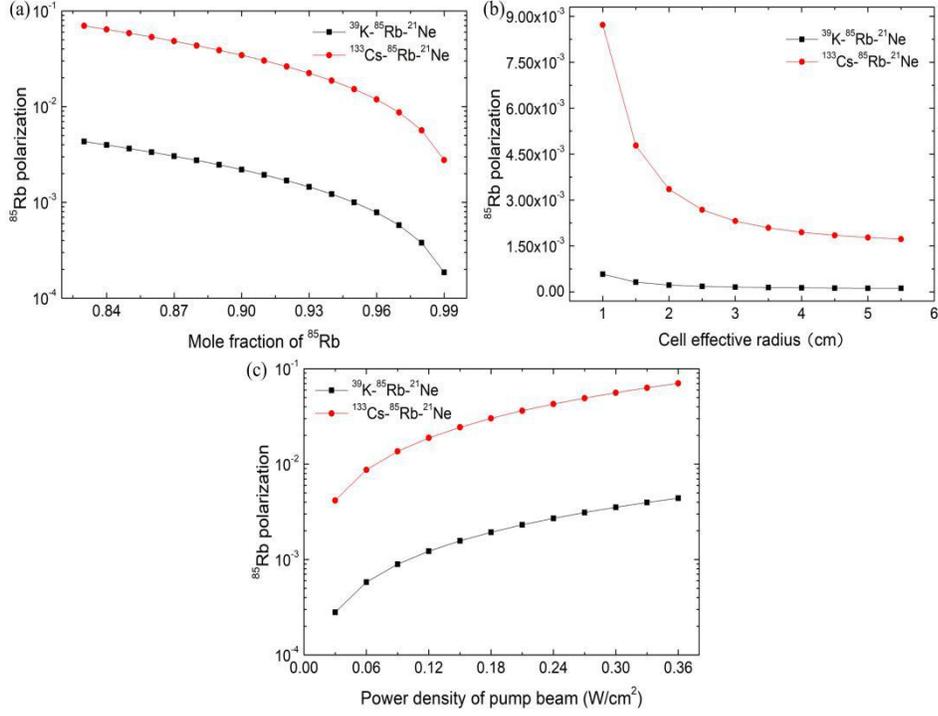

Fig.3. The $^{85}$Rb polarization of $^{39}$K ($^{133}$Cs)-$^{85}$Rb-$^{21}$Ne co-magnetometers versus the mole fraction of $^{85}$Rb in (a), the cell effective radius in (b) and the power density of pump beam in (c).

Fig. 3 shows that the $^{85}$Rb polarization of $^{39}$K ($^{133}$Cs)-$^{85}$Rb-$^{21}$Ne co-magnetometers changes with the mole fraction of $^{85}$Rb $f_{Rb}$, cell effective radius $a$ and power density of pump beam $I_{PI}$.

Fig. 3(a) shows that $^{85}$Rb polarizations decrease with increasing mole fraction of $^{85}$Rb $f_{Rb}$. From the formulae of the $D_r$, $R_{SD}^{Rb}$, $B^e$ and alkali-alkali spin exchange collisions relaxation rate, we can find that $D_r$, $B^e$ and alkali-alkali spin exchange collisions relaxation rate decrease and $R_{SD}^{Rb}$ increases with the increasing $f_{Rb}$. Therefore, the decrease process of the $^{85}$Rb polarization is determined by $D_r$, $R_{SD}^{Rb}$, $B^e$ and $R_{KRbNe}^{RbK}$, $R_{CsRbNe}^{RbCs}$, $R_{KRbNe}^{RbRb}$, $R_{CsRbNe}^{RbRb}$, $R_{KRbNe}^{EE}$, $R_{CsRbNe}^{EE}$, where $D_r$ and $R_{SD}^{Rb}$ play important roles.

Figure 3(b) demonstrates that $^{85}$Rb polarizations decrease with increasing cell effective radius $a$. From the formula of the $^{85}$Rb polarization, $R_{Wall}$ and $B^e$, we can find that the wall relaxation rate $R_{wall-K-Rb-Ne}^{Rb-N_2}$, $R_{wall-Cs-Rb-Ne}^{Rb-N_2}$, $R_{wall-K-Rb-Ne}^{Rb-Ne-N_2}$, $R_{wall-Cs-Rb-Ne}^{Rb-Ne-N_2}$, $R_{wall-K-Rb-Ne}^{Rb-Ne}$ and $R_{wall-Cs-Rb-Ne}^{Rb-Ne}$ decrease when $a$ increases in Fig. 1(c) and $B^e$ increases with increasing $a$. Therefore, the decrease process of the $^{85}$Rb polarization is determined by $R_{Wall}$ and $B^e$, where the wall relaxation rate plays a decisive role.

Fig. 3(c) shows that $^{85}$Rb polarization increases when $I_{PI}^{K}$ ($I_{PI}^{Cs}$) increases. From the formula of $\Gamma_{D1}^{K}$ ($\Gamma_{D1}^{Cs}$), $R_p$, frequency shift and $B^e$, we can find that $\Gamma_{D1}^{K}$ ($\Gamma_{D1}^{Cs}$) increases with the increasing $I_{PI}^{K}$ ($I_{PI}^{Cs}$), which lead to $R_p$, frequency shift and $B^e$ increase with increasing $I_{PI}^{K}$ ($I_{PI}^{Cs}$). $R_p$ plays a decisive role for the change of $^{85}$Rb polarization with $I_{PI}^{K}$ ($I_{PI}^{Cs}$). The $^{85}$Rb polarization in the $^{39}$K-$^{85}$Rb-$^{21}$Ne co-magnetometer is smaller than the one in the $^{133}$Cs-$^{85}$Rb-$^{21}$Ne co-magnetometer.

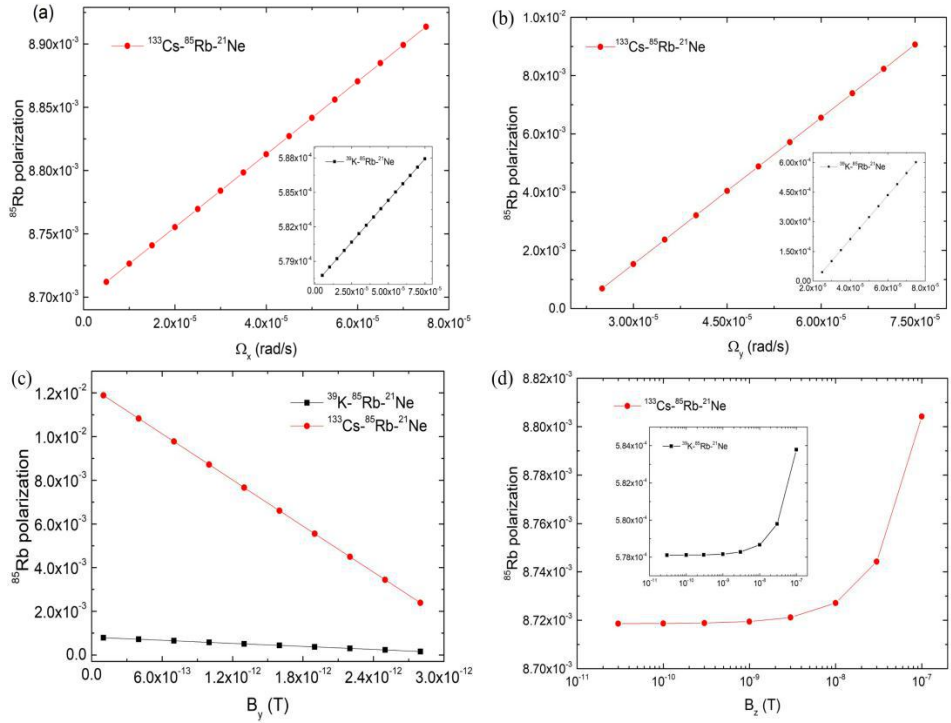

Fig.4. The $^{85}$Rb polarization of $^{39}$K ($^{133}$Cs)-$^{85}$Rb-$^{21}$Ne co-magnetometers versus the input rotation velocity in the $x$ direction in (a) and $y$ direction in (b), the external magnetic field in the $y$ direction $B_y$ in (c) and $z$ direction $B_z$ in (d), respectively.

Fig. 4(a) and Fig. 4(b) demonstrate that the $^{85}$Rb polarizations increase with increasing input rotation velocity in the $x$ direction $\Omega_x$ and the $y$ direction $\Omega_y$, respectively. $\Omega/\gamma^n$ is equivalent to a magnetic field, which can be compensated by the compensation magnetic field $\delta B_z$. When $L_z$ is some negative value, $\dfrac{\Omega_y}{\gamma^n}+\dfrac{L_z}{R_{tot}^{Rb}/\gamma_e}\dfrac{\Omega_x}{\gamma^n}=0$, which can help to eliminate the effect of the input rotation velocity $\Omega_x$ and $\Omega_y$ on the $^{85}$Rb polarization. Fig. 4(c) shows that $^{85}$Rb polarizations decrease with increasing external magnetic field in the $y$ direction $B_y$. a $B_y$ modulation method could be utilized to measure the mixed ac Stark shifts[43]. Figure 4(d) demonstrates that $^{85}$Rb polarizations increase with increasing external magnetic field in the $z$ direction $B_z$. The change of $B_z$ will make the compensation magnetic field $\delta B_z$ change, which will affect the $^{85}$Rb polarization.

### 3.3. The calculation result of the fundamental sensitivity

We take $^{39}$K($^{133}$Cs) as A atom, $^{85}$Rb as B atom, one of the $f_{Rb}$, $n_{Ne}$, $n_{N2}$, $T$, $\lambda_K$ ($\lambda_{Cs}$), $I_{PI}^K$ ($I_{PI}^{Cs}$), $a$, $B$ (it is helpful for us to study how $B$ influence the SERF regime and fundamental sensitivity of the co-magnetometers) and $V$ in Eq. (12) as a variable (other parameters are invariable) to obtain the results that the fundamental sensitivity of the co-magnetometers based on $^{39}$K-$^{85}$Rb-$^{21}$Ne and $^{133}$Cs-$^{85}$Rb-$^{21}$Ne change with the variable. Depending on suggestions and the typical conditions of the experiment group,[34, 35, 54] in order to facilitate the theoretical analysis, we take $n_{Ne}$= 2×10$^{19}$ cm$^{-3}$, $n_{N2}$ =2×10$^{17}$ cm$^{-3}$, $\lambda_K$=769.808 nm, $\lambda_{Cs}$=894.3 nm, $I_{PI}^K = I_{PI}^{Cs}$ =0.06W/cm$^2$, $B$=10$^{-12}$ T ( $R_{SE}^{ee} \approx 0$), $T$=457.5 K, $a$ =1 cm, $a_p$ = 1 cm, $f_{Rb}$=0.97, $V$=1 cm$^3$ and $t$=1s.

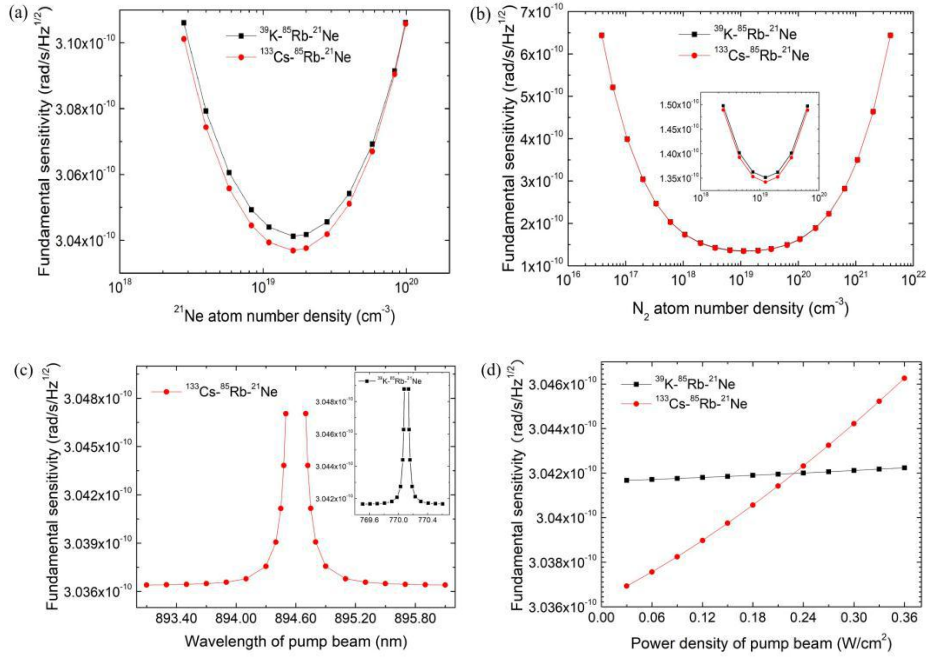

Fig.5. The fundamental sensitivity of $^{39}$K ($^{133}$Cs)-$^{85}$Rb-$^{21}$Ne co-magnetometers versus the number density of the buffer gas $^{21}$Ne in (a), number density of the quench gas N$_2$ in (b), the wavelength of pump beam in (c) and the power density of pump beam in (d).

Figure 5(a) represents that the fundamental sensitivity of $^{39}$K ($^{133}$Cs)-$^{85}$Rb-$^{21}$Ne co-magnetometer augments with the increasing number density of $^{21}$Ne $n_{Ne}$ when $n_{Ne}$ is smaller than a critical value about $1.6230\times10^{19}$ cm$^{-3}$ and decreases when $n_{Ne}$ is larger than the value. For this phenomenon, just as we discussed in our previous work,[24] we think that more alkali-metal atoms diffuse to the cell wall and less spin exchange collisions between alkali-metal atoms A and B when $n_{Ne}$ is smaller than the value and decrease. Less alkali-metal atoms diffuse to the cell wall and more spin exchange collisions between alkali-metal atoms and buffer gas so that there are less spin exchange collisions in alkali-metal atoms when $n_{Ne}$ is larger than the value and increase. Hence, if we take the critical value as $n_{Ne}$, spin exchange collisions in alkali-metal atoms are the most, we can obtain the highest fundamental sensitivity of the co-magnetometer. From Eq. (12), Fig. 1(a) and $R_p$, we can find that $R_{SD}$ increases with the increasing $n_{Ne}$, $R_{wall-K-Rb-Ne}^{Rb-Ne}$ and $R_{wall-Cs-Rb-Ne}^{Rb-Ne}$ decrease with increasing $n_{Ne}$, $R_p^K$ and $R_p^{Cs}$ increase with increasing $n_{Ne}$ when $n_{Ne}$ increases, other systemic parameters in Eq. (12) do not change with $n_{Ne}$, hence, there is a smallest $n_{Ne}$ to make the fundamental sensitivity highest, which is determined by $R_{SD}$, $R_{wall-K-Rb-Ne}^{Rb-Ne}$ ($R_{wall-Cs-Rb-Ne}^{Rb-Ne}$) and $R_p^K$ ($R_p^{Cs}$).

Figure 5(b) shows that the fundamental sensitivity of $^{39}$K($^{133}$Cs)-$^{85}$Rb-$^{21}$Ne co-magnetometer increases with the increasing N$_2$ number density $n_{N2}$ when $n_{N2}$ is smaller than a critical value about $1.2407\times10^{19}$ cm$^{-3}$ and decreases when $n_{N2}$ is higher than the value. From Eq. (12) and Fig.1(b), we can find that $R_{SD}$ increases with the increasing $n_{N2}$, $R_{wall-K-Rb-Ne}^{Rb-Ne-N_2}$ and $R_{wall-Cs-Rb-Ne}^{Rb-Ne-N_2}$ decrease with increasing $n_{N2}$, other systemic parameters in Eq. (12) do not change with $n_{N2}$, hence, there is a smallest $n_{N2}$ to make the fundamental sensitivity highest, which is determined by $R_{SD}$ and $R_{wall-K-Rb-Ne}^{Rb-N_2}$ ($R_{wall-Cs-Rb-Ne}^{Rb-N_2}$).

Figure 5(c) demonstrates that the fundamental sensitivity of $^{39}$K($^{133}$Cs)-$^{85}$Rb-$^{21}$Ne co-magnetometers decrease when the wavelength of pump beam $\lambda_K$($\lambda_{Cs}$) increases but is smaller than the value of about 770.085 nm (894.5032 nm) and increase when $\lambda_K(\lambda_{Cs})$ is larger than the value of about 770.1309 nm (894.696 nm)($^{85}$Rb polarization is about 0.08).

Figure 5(d) shows that the fundamental sensitivity decreases with the increasing power density of pump beam $I_{PI}^K$ ($I_{PI}^{Cs}$). In the experiment, $I_{PI}$ is usually a few hundreds of mW/cm$^2$. Hence $\lambda_K(\lambda_{Cs})$ and $I_{PI}^K$ ($I_{PI}^{Cs}$) affect the fundamental sensitivity by changing $R_p^K$ ($R_p^{Cs}$).

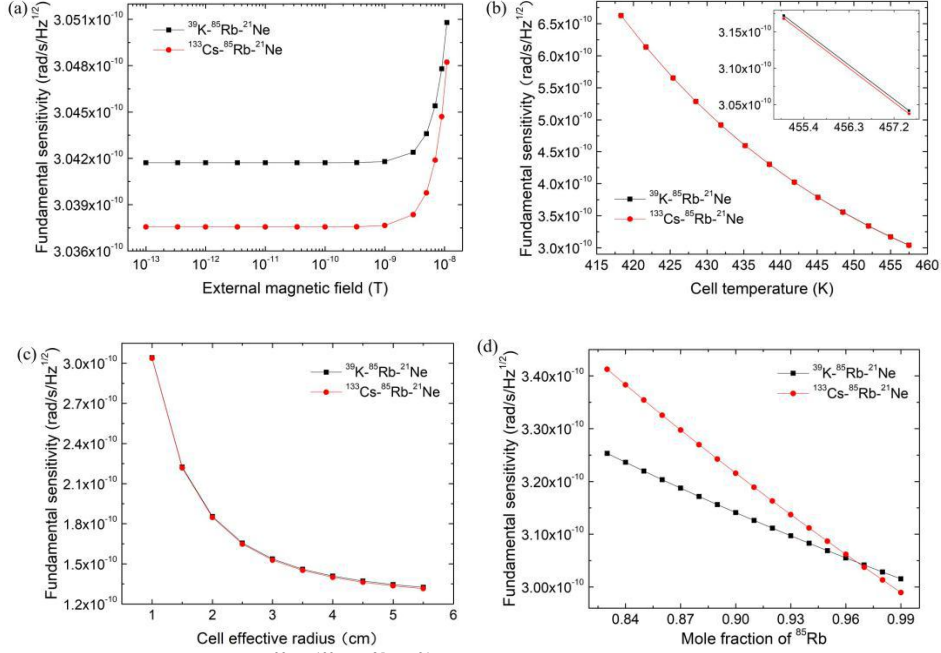

Fig.6. The fundamental sensitivity of $^{39}$K ($^{133}$Cs)-$^{85}$Rb-$^{21}$Ne co-magnetometers change with the external magnetic field in (a), the cell temperature in (b), the cell effective radius in (c) and mole fraction of $^{85}$Rb in (d).

Figure 6(a) describes that the fundamental sensitivity of $^{39}$K ($^{133}$Cs)-$^{85}$Rb-$^{21}$Ne co-magnetometers decrease with the increasing external magnetic field $B$. The increasing $B$ will make the relaxation rate due to alkali-alkali spin-exchange collisions $R_{SE}^{EE}$ increase, which makes the fundamental sensitivity decrease. The fundamental sensitivity of $^{39}$K ($^{133}$Cs)-$^{85}$Rb-$^{21}$Ne co-magnetometers increase with the increasing cell temperature $T$ in Fig. 6(b) and we can know that the fundamental sensitivity of $^{39}$K-$^{85}$Rb-$^{21}$Ne co-magnetometer is lower than the one of $^{133}$Cs-$^{85}$Rb-$^{21}$Ne co-magnetometer from the illustration. From Eq. (12), Fig. 1(c), the formula of $R_p$ and alkali-alkali spin exchange collision relaxation rate, we can find that $R_{SD}$ and $n_B$ ($n_{Rb}$) increase with the increasing $T$, $R_{wall}$ increases rapidly when the $T$ increases and $R_p$ increases slowly with the increasing $T$. However, there is a larger increase for the $^{85}$Rb number density and a decrease for $R_{SE}^{EE}$ ($R_{SE}^{EE} \approx 0$) when the $T$ increases. Hence, the fundamental sensitivity increases with the increasing $T$, which is mainly determined by the $^{85}$Rb number density.

The fundamental sensitivity of $^{39}$K ($^{133}$Cs)-$^{85}$Rb-$^{21}$Ne co-magnetometers increase with the increasing cell effective radius $a$, respectively in Fig. 6(c) duo to the fact that the increasing $a$ makes the $R_{wall}$ decrease, which will increase the fundamental sensitivity. The fundamental sensitivity of $^{39}$K ($^{133}$Cs)-$^{85}$Rb-$^{21}$Ne co-magnetometers does not change with increasing $a_p$. From the formula of $R_p$, we can find that $R_p$ does not change with the increasing $a_p$ duo to the fact that $a_p$ in $\Phi_{D1}^K$ ($\Phi_{D1}^{Cs}$) and $A_t$ are canceled out.

Figure 6(d) describes that the fundamental sensitivity increases with the increasing mole fraction of $^{85}$Rb $f_{Rb}$, duo to the alkali-alkali spin exchange collisions relaxation rate decreases and $^{85}$Rb number density increases with the increasing $f_{Rb}$. When the $f_{Rb}$ is larger than about 0.9662, the fundamental sensitivity of $^{133}$Cs-$^{85}$Rb-$^{21}$Ne co-magnetometer is higher than the one of $^{39}$K-$^{85}$Rb-$^{21}$Ne co-magnetometer. From Eq. (12), the fundamental sensitivity of $^{39}$K ($^{133}$Cs)-$^{85}$Rb-$^{21}$Ne co-magnetometers increases with increasing measurement volume.

As a result, the polarization of the $^{85}$Rb atom of the co-magnetometer based on $^{133}$Cs-$^{85}$Rb-$^{21}$Ne is bigger than the one based on $^{39}$K-$^{85}$Rb-$^{21}$Ne in Figs. 2-4. The fundamental sensitivity of $^{133}$Cs-$^{85}$Rb-$^{21}$Ne co-magnetometer is higher than the $^{39}$K-$^{85}$Rb-$^{21}$Ne co-magnetometer at the same cell temperature in the SERF regime with the same frequency detuning of the pump beam when the external magnetic field is smaller than about $1.7884\times10^{-8}$ T, the mole fraction of $^{85}$Rb is bigger than about 0.9662, or the power density of pump beam is smaller than about 0.229 W/cm$^2$ in Fig. 5 and 6 under our chosen conditions.

We obtain a fundamental sensitivity of about $7.542\times10^{-11}$ rad/s/Hz$^{1/2}$ with $^{39}$K-$^{85}$Rb-$^{21}$Ne co-magnetometer with $^{85}$Rb polarization is about $4.4892\times10^{-5}$ and $n_K/n_{Rb} \approx 5.1228\times10^{-4}$, a fundamental sensitivity of about $7.5355\times10^{-11}$ rad/s/Hz$^{1/2}$ with $^{133}$Cs-$^{85}$Rb-$^{21}$Ne co-magnetometer with $^{85}$Rb polarization is about $8.193\times10^{-4}$ and $n_{Cs}/n_{Rb} \approx 0.0043$ when $n_{Ne} = 6\times10^{19}$ cm$^{-3}$, $n_{N2} = 6.3\times10^{18}$ cm$^{-3}$, $T$ = 406.696 K, $f_{Rb}$= 0.99, $I_{PI}^K = I_{PI}^{Cs} = 0.06$ W/cm$^2$, $\lambda_K$= 769.938 nm, $\lambda_{Cs}$=894.43 nm, $a$ = 2 cm, $a_p$ = 1 cm, $V$ = 10 cm$^3$, $B = 10^{-12}$ T and $t$ = 1 s. The fundamental sensitivity of $^{39}$K($^{133}$Cs)-$^{85}$Rb-$^{21}$Ne co-magnetometer is higher than the angular velocity sensitivity of $2.1\times10^{-8}$ rad/s/Hz$^{1/2}$ of K-Rb-$^{21}$Ne comagnetometer[61]. By optimizing the above parameters, we obtain a fundamental sensitivity of about $3.9993 \times 10^{-11}$ rad/s/Hz$^{1/2}$ with $^{39}$K-$^{85}$Rb-$^{21}$Ne with the polarization of the $^{85}$Rb atom is about 0.0021 and $n_K/n_{Rb} \approx 7.5791 \times 10^{-4}$, a fundamental sensitivity of $3.8221\times10^{-11}$ rad/s/Hz$^{1/2}$ with $^{133}$Cs-$^{85}$Rb-$^{21}$Ne with the polarization of the $^{85}$Rb atom is about 0.0224 and $n_{Cs}/n_{Rb} \approx 0.0049$ with $n_{Ne} = 8.3407\times10^{18}$ cm$^{-3}$, $n_{N2} = 6.2035\times10^{18}$ cm$^{-3}$, $T$ = 457.5 K, $f_{Rb}$ = 0.99, $I_{PI}^K = I_{PI}^{Cs} = 0.06$ W/cm$^2$, $\lambda_K$= 770.038 nm, $\lambda_{Cs}$=894.53 nm, $a$ = 2 cm, $a_p$ = 1 cm, $V$ = 10 cm$^3$, $B = 10^{-12}$ T and $t$ = 1 s, with higher fundamental sensitivity possible at larger measurement volume, proper amount of buffer gas and quench gas, smaller pumping rate of pump beam and higher temperature.

4. **Conclusion**

In conclusion, we find that the $^{85}$Rb polarization of $^{133}$Cs-$^{85}$Rb-$^{21}$Ne co-magnetometer is larger than the one of $^{39}$K-$^{85}$Rb-$^{21}$Ne co-magnetometer in our chosen conditions. The fundamental sensitivity of $^{133}$Cs-$^{85}$Rb-$^{21}$Ne co-magnetometer is higher than the one of $^{39}$K-$^{85}$Rb-$^{21}$Ne co-magnetometer when the external magnetic field is smaller than about $1.7884\times10^{-8}$ T, the mole fraction of $^{85}$Rb is larger than about 0.9662 or the power density of pump beam is smaller than about 0.229 W/cm$^2$.

To obtain a higher fundamental sensitivity between $^{39}$K-$^{85}$Rb-$^{21}$Ne and $^{133}$Cs-$^{85}$Rb-$^{21}$Ne co-magnetometers, we should choose $^{133}$Cs-$^{85}$Rb-$^{21}$Ne co-magnetometer (when the external magnetic field is smaller than about $1.7884\times10^{-8}$ T, the mole fraction of $^{85}$Rb is larger than about 0.9662 or the power density of pump beam is smaller than about 0.229 W/cm$^2$) with $^{21}$Ne atoms as the buffer gas, take the critical values of $^{21}$Ne number density and quench gas N$_2$ number density, increase the cell effective radius, the measurement volume, the cell temperature (when the quantity of alkali metal atoms are enough) and mole fraction of $^{85}$Rb atoms, reduce the external magnetic field and power density of pump beam, choose suitable wavelength of pump beam based on actual demand of the fundamental sensitivity and spatial resolution. We estimate the fundamental sensitivity limit of the co-magnetometers due to the shot noise superior to $7.5355\times10^{-11}$ rad/s/Hz$^{1/2}$ with $^{133}$Cs-$^{85}$Rb-$^{21}$Ne co-magnetometer, which is higher than the one of a K-$^{21}$Ne atomic co-magnetometer of $2\times10^{-10}$ rad/s/Hz$^{1/2}$. We could choose suitable conditions on the basis of the experiment requirements to gain a higher sensitivity of the co-magnetometers, keep the costs down and carry forward the miniaturization and practical application of the co-magnetometers.

Table 1. Parameters used for the calculation.

| Parameter | Value |
|---|---|
| Boltzmann's constant $k_B$ | $1.38\times10^{-23}$ J/K |
| Atomic mass unit $m$ | $1.660539040(20)\times10^{-27}$ kg |
| $\pi$ | 3.14 |
| $^{21}$Ne nuclear gyromagnetic ratio[59] | $2.1\times10^{7}$ rad/s/T |
| Electron spin $g$ factor | $2\times1.001159657$ |
| Planck's constant $\hbar$ | $1.054589\times10^{-34}$ J·s |
| Bohr magneton $\mu_B$ | $9.27408\times10^{-24}$ J/T |
| $\sigma_{SE}^{K\ [55]}$ | $1.8\times10^{-14}$ cm$^2$ |
| $\sigma_{SE}^{Rb\ [52]}$ | $1.9\times10^{-14}$ cm$^2$ |
| $\sigma_{SE}^{Cs\ [52]}$ | $2.1\times10^{-14}$ cm$^2$ |
| $\sigma_{SD}^{K\ [62]}$ | $1\times10^{-18}$ cm$^2$ |
| $\sigma_{SD}^{Rb\ [46]}$ | $1.6\times10^{-17}$ cm$^2$ |
| $\sigma_{SD}^{Cs\ [63]}$ | $2\times10^{-16}$ cm$^2$ |
| $\sigma_{SD}^{K\text{-}Ne\ [36,39]}$ | $1\times10^{-23}$ cm$^2$ |
| $\sigma_{SD}^{Rb\text{-}Ne\ [56]}$ | $5.2\times10^{-23}$ cm$^2$ |
| $\sigma_{SD}^{Cs\text{-}Ne\ [57]}$ | $1\times10^{-22}$ cm$^2$ |
| $\sigma_{SD}^{K\text{-}N_2\ [27,58]}$ | $7.9\times10^{-23}$ cm$^2$ |
| $\sigma_{SD}^{Rb\text{-}N_2\ [27,36]}$ | $1\times10^{-22}$ cm$^2$ |
| $\sigma_{SD}^{Cs\text{-}N_2\ [27,36]}$ | $5.5\times10^{-22}$ cm$^2$ |

This work was supported by the National Key R&D Program of China under grants No. 2016YFA0301500, NSFC under grants Nos. 61835013, Strategic Priority Research Program of the Chinese Academy of Sciences under grants Nos. XDB01020300, XDB21030300.

**References**

1. Schreiber K U, Klügel T, Wells J P, Hurst R B and Gebauer A 2011 Phys. Rev. Lett. **107**, 173904
2. Schreiber K U and Wells J P R 2013 Rev. Sci. Instrum. **84**, 041101
3. Shahriar M S and Salit M 2008 J. Mod. Opt. **55**, 3133
4. Stedman G E1997 Rep. Prog. Phys. **60**, 615
5. Everitt C W F, DeBra D B, Parkinson B W, et al 2011 Phys. Rev. Lett. **106**, 221101
6. Lefevre H C 2013 Opt. Fiber. Technol. **19**, 828
7. Kominis I K, Kornack T W, Allred J C and Romalis M V 2003 Nature **422**, 596
8. Kornack T W, Ghosh R K and Romalis M V 2005 Phys. Rev. Lett. **95**, 230801
9. Meyer D and Larsen M 2014 Gyroscopy Navigation **5**, 75
10. Brown J M, Smullin S J, Kornack T W, Romalis M V 2010 Phys. Rev. Lett. **105**, 151604
11. Smiciklas M, Brown J M, Cheuk L W, Smullin S J and Romalis M V 2011 Phys. Rev. Lett. **107**, 171604
12. Vasilakis G, Brown J M, Kornack T W and Romalis M V 2009 Phys. Rev. Lett. **103**, 261801
13. Tullney K, Allmendinger F, Burghoff M, Heil W, Karpuk S, Kilian W, Knappe-Grüneberg S, Müller W, Schmidt U, Schnabel A, Seifert F, Sobolev Y and Trahms L 2013 Phys. Rev. Lett. **111**, 100801
14. Bulatowicz M, Griffith R, Larsen M, Mirijanian J, Walker T G, Fu C B, Smith E, Snow W M and Yan H 2013 Phys. Rev. Lett. **111**, 102001
15. Arvanitaki A and Geraci A A 2014 Phys. Rev. Lett. **113**, 161801
16. Rosenberry M A and Chupp T E 2001 Phys. Rev. Lett. **86**, 22
17. Kornack T W 2005 PhD Dissertation (New Jersey: Princeton University)
18. Fang J C, Wan S A, Qin J, Zhang C, Quan W, Yuan H and Dong H F 2013 Rev. Sci. Instrum. **84**, 083108

19. Zeng X, Wu Z, Call T, Miron E, Schreiber D and Happer W 1985 Phys. Rev. A **31**, 260
20. Jiang L W, Quan W, Li R J, Duan L H, Fan W F, Wang Z, Liu F, Xing L and Fang J C 2017 Phys. Rev. A **95**, 062103
21. Jiang L W, Quan W, Li R J, Fan W F and Fang J C 2018 Appl. Phys. Lett. **112**, 054103
22. Quan W, Wei K, Zhao T, Li H R, Zhai Y Y 2019 Phys. Rev. A. **100**, 012118
23. Jiang LW, Quan W, Liu F, Fan W F, Xing L, Duan L H, Liu W M and Fang J C 2019 Phys. Rev. Applied. **12**, 024017
24. Liu J H, Jing D Y, Wang L L, Li Y, Quan W, Fang J C and Liu W M 2017 Sci. Rep. **7**, 6776
25. Alcock C B, Itkin V P and Horrigan M K 1984 Can. Metall. Quart. **23**, 309
26. Ito Y, Sato D, Kamada K and Kobayashi T 2016 Opt. Express **24**, 015391
27. Seltzer S J 2008 PhD Dissertation (New Jersey: Princeton University)
28. Happer W and Tam A C 1977 Phys. Rev. A **16**, 1877
29. Dong H F, Xuan L F, Zhuo C and Lin H B 2012 J. Test Measurement Technol. **26**, 468(in Chinese)
30. Babcock E, Nelson I, Kadlecek S, Driehuys B, Anderson L W, Hersman F W and Walker T G 2003 Phys. Rev. Lett. **91**, 123003
31. Romalis M V, Miron E and Cates G D 1997 Phys. Rev. A **56**, 4569
32. Citron M L, Gray H R, Gabel C W and Stroud C R 1977 Phys. Rev. A **16**, 1507
33. Cole H R and Olson R E 1985 Phys. Rev. A **31**, 2137
34. Fang J C, Wang T, Zhang H, Li Y and Zou S 2014 Rev. Sci. Instrum. **85**, 123104
35. Chen Y, Quan W, Duan L H, Lu Y, Jiang L W and Fang J C 2016 Phys. Rev. A **94**, 052705
36. Allred J C, Lyman R N, Kornack T W and Romalis M V 2002 Phys. Rev. Lett. **89**, 130801
37. Schaefer S R, Cates G D, Chien T R, Gonatas D, Happer W and Walker T G 1989 Phys. Rev. A **39**, 5613
38. Ghosh R K and Romalis M V 2010 Phys. Rev. A **81**, 043415
39. Franz F A and Volk C 1982 Phys. Rev. A **26**, 85
40. Franz F A and Sooriamoorthi C E 1974 Phys. Rev. A **10**, 126
41. Franz F A and Volk C 1976 Phys. Rev. A **14**, 1711
42. Silver J A 1984 J. Chem. Phys. **81**, 5125
43. Fang J C, Chen Y, Zou S, Liu X J, Hu Z H, Quan W, Yuan H, Ding M 2016 J. Phys. B **49**, 065006
44. Ledbetter M P, Savukov I M, Acosta V M, Budker D and Romalis M V 2008 Phys. Rev. A **77**, 033408
45. Hager G D, Lott G E, Archibald A J, Blank L, Weeks D E and Perram G P 2014 J. Quant. Spectrosc. Radiat. Transf. **147**, 261
46. Walker T G and Happer W 1997 Rev. Mod. Phys. **69**, 629
47. Migdalek J and Kim Y K 1998 J. Phys. B **31**, 1947
48. Caliebe E and Niemax K 1979 J. Phys. B **12**, L45
49. Lwin N and McCartan D G 1978 J. Phys. B **11**, 3841
50. Appelt A, Baranga B A, Erickson C J, Romalis M V, Young A R and Happer W 1998 Phys. Rev. A **58**, 1412
51. Gibbs H M and Hull R J 1967 Phys. Rev. **153**, 132
52. Ressler N W, Sands R H and Stark T E 1969 Phys. Rev. **184**, 102
53. Shao W J, Wang G D and Hughes E W 2005 Phys. Rev. A **72**, 022713
54. Quan W, Li Y and Liu B 2015 Europhys. Lett. **110**, 60002
55. Aleksandrov E B, Balabas M V, Vershovskii A K, Okunevich A I and Yakobson N N 1999 Opt. Spectrosc. **87**, 329
56. Franzen W 1959 Phys. Rev. **115**, 850
57. Beverini N, Minguzzi P and Strumia F 1972 Phys. Rev. A **5**, 993
58. Kadlecek S, Anderson L W and Walker T 1998 Nucl. Instrum. Methods. Phys. Res. **402**, 208
59. Fang J C and Qin J 2012 Sensors **12**, 6331
60. Fang J C, Li R J, Duan L H, Chen Y and Quan W 2015 Rev. Sci. Instrum. **86**, 073116
61. Chen Y, Quan W, Zou S, Lu Y, Duan L H, Li Y, Zhang H, Ding M and Fang J C 2016 Sci. Rep. **6**, 36547
62. Vliegen E, Kadlecek S, Anderson L W, Walker T G, Erickson C J and Happer W 2001 Nucl. Instrum. Methods Phys. Res. A **460**, 444
63. Bhaskar N D, Pietras J, Camparo J, Happer W and Liran J 1980 Phys. Rev. Lett. **44**, 930